# ISEGEN: Generation of High-Quality Instruction Set Extensions by Iterative Improvement *

Partha Biswas         Sudarshan Banerjee         Nikil Dutt
partha@cecs.uci.edu   banerjee@cecs.uci.edu      dutt@cecs.uci.edu

Center for Embedded Computer Systems
Donald Bren School of Information and Computer Science
University of California, Irvine
Irvine, CA 92697-3425, USA

Laura Pozzi                    Paolo Ienne
laura.pozzi@epfl.ch            paolo.ienne@epfl.ch

Ecole Polytechnique Fédérale de Lausanne (EPFL)
School of Computer and Communication Sciences
CH-1015 Lausanne, Switzerland

## Abstract

*Customization of processor architectures through Instruction Set Extensions (ISEs) is an effective way to meet the growing performance demands of embedded applications. A high-quality ISE generation approach needs to obtain results close to those achieved by experienced designers, particularly for complex applications that exhibit regularity: expert designers are able to exploit manually such regularity in the data flow graphs to generate high-quality ISEs. In this paper, we present ISEGEN, an approach that identifies high-quality ISEs by iterative improvement following the basic principles of the well-known* Kernighan-Lin (K-L) *min-cut heuristic. Experimental results on a number of MediaBench, EEMBC and cryptographic applications show that our approach matches the quality of the optimal solution obtained by exhaustive search. We also show that our ISEGEN technique is on average $20\times$ faster than a genetic formulation that generates equivalent solutions. Furthermore, the ISEs identified by our technique exhibit $35\%$ more speedup than the genetic solution on a large cryptographic application (AES) by effectively exploiting its regular structure.*

## 1 Introduction

Continuing advances in manufacturing processes have made it possible for processor vendors to build increasingly fast processors. However, newer applications place an increasing demand on performance, at a rate faster than that achievable by processors. These trends have necessitated the migration of critical computations from the processor core to an application-specific unit that is able to perform compute-intensive tasks efficiently. We call such a unit an *Ad-hoc Functional Unit (AFU)*. The AFU accelerates critical operations of application algorithms by executing application-specific *Instruction Set Extensions (ISEs)*.

Automatic generation of ISEs is essentially the task of hardware-software partitioning applied at an instruction-level granularity. The Kernighan-Lin (K-L) min-cut algorithm is a well-known graph partitioning heuristic originally designed for circuit partitioning [2]. Recently, this heuristic has been successfully adapted for task-level partitioning of a system into hardware and software [1]. In this paper, we apply the K-L heuristic at the instruction-level granularity to automatically generate ISEs. We refer to our approach as ISEGEN. Our motivation for employing an iterative improvement technique like K-L is to generate solutions close to those obtained manually by expert designers. In order to match such a solution quality, the control parameters of ISEGEN closely model the decisions taken by the designer.

We show the efficacy of ISEGEN on a number of em-

*This work was partially supported by NSF grants: CCR-0203813, CCR-0205712 and SRC contract: 2003-HJ1111.

bedded applications selected from MediaBench, EEMBC and cryptographic suites by comparing our results with the best known approaches of ISE generation. We demonstrate that ISEGEN runs up to $29\times$ faster than a previous genetic formulation while yielding ISEs having speedup comparable with the optimal solution [3]. On a large cryptographic application (AES) for which the exhaustive techniques fail, ISEGEN — by effectively exploiting its regular structure — generates $35\%$ more speedup than the genetic approach [4].

The rest of the paper is organized as follows. In Section 2, we define our problem. In Section 3, we discuss related research work and our motivation. We propose our ISEGEN approach in Section 4. In Section 5, we describe the experimental results that demonstrate the efficacy of our approach. Finally, Section 6 concludes the paper.

## 2 Problem Definition

Instructions within a basic block are typically represented as a Directed Acyclic Graph (DAG), $G = (V, E)$: the nodes $V$ represent instructions and the edges $E$ capture the data dependencies between them. We define a *cut* $C$ representing a potential ISE as a subgraph of $G$, $C \subseteq G$ Let $\mathrm{M}(C)$ be the function that measures the merit of a cut $C$ as an estimation of the speedup achievable by implementing $C$ as an ISE. Let $I_{\mathrm{ISE}}(C)$ and $O_{\mathrm{ISE}}(C)$ respectively be the number of inputs and the number of outputs of $C$. The maximum number of operands of an ISE (or a cut) is limited by the number of register file ports in the underlying core.

Let $N_{in}$ and $N_{out}$ be the maximum number of input and output operands respectively. A cut $C$ is architecturally feasible if its inputs are available at the time of issue. This is only possible if $C$ is convex, i.e., if there exists no path from a node $u \in C$ to another node $v \in C$ through a node $w \notin C$ [3]. The problem of ISE generation can be broken into the following two sub-problems:

**Problem 1** *Given the data flow graph (DFG) $G = (V, E)$ in a basic block, find a cut $C \subseteq G$ that maximizes $\mathrm{M}(C)$ under the following constraints:*

- *Input-Output (I/O) Constraints: $I_{\mathrm{ISE}}(C) \leq N_{\mathrm{in}}$ and $O_{\mathrm{ISE}}(C) \leq N_{\mathrm{out}}$.*
- *Convexity Constraint: $C$ is convex.*

**Problem 2** *Given the basic blocks in an application and the maximum allowed number of ISEs as $N_{\mathrm{ISE}}$, find cuts that maximize the speedup achievable for the entire application.*

## 3 State of the Art and Motivation

Some of the earlier work in ISE generation applied to reconfigurable computing [11, 10] considers only single-output subgraphs in ISE generation. Even though a few recently proposed approaches [5, 6] handle multiple outputs, they identify only connected subgraphs. However, the opportunity to include independent subgraphs in the same ISE exposes speedup potentials, while algorithms identifying only connected graphs are unable to exploit high constraints of ISE outputs. Therefore, we also consider independent subgraphs in ISEGEN.

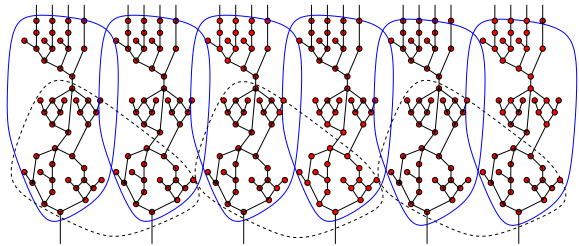

**Figure 1. An example showing the advantage of large scale reuse — Finding three instances of the largest ISE (shown with a dotted boundary) is not as effective as finding a large ISE with six instances (shown with a solid boundary).**

When the goal of ISE generation is speedup coupled with dynamic reuse, as in [7, 8], the resulting subgraphs are generally small. In practice, if one wants to mimic the excellent results targeted by expert designers, clusters of 2 or 3 instructions are far too small for arousing real interest: typical results at this level generally include only peculiar address generation patterns, pre- or post-shifting, or well-known arithmetic patterns such as multiply-accumulators. There is a need for algorithms that can identify large *and* reusable clusters, efficiently covering the application DFG. Figure 1 demonstrates this principle with the help of an example. This motivates our ISEGEN approach that not only generates ISEs having higher potential for speedup, but which also shows the efficacy of the generated ISEs in terms of their reusability.

An exact solution [3] that uses an exhaustive search with pruning is not practical for applications having large basic-blocks. A genetic formulation [4] presents a practical solution with results showing good speedup for the generated ISEs. However, the genetic algorithm is stochastic in nature and therefore multiple runs may result in different solutions. Our ISEGEN approach, on the other hand, is an iterative improvement technique that closely mimics the decisions taken by an expert designer; consequently we are able to match the solution quality of expert designers.

## 4 The ISEGEN Approach

We reiterate that ISEGEN essentially performs Hardware-Software partitioning at instruction-level granularity. The instructions belonging to the hardware

Proceedings of the Design, Automation and Test in Europe Conference and Exhibition (DATE'05)
1530-1591/05 $ 20.00 IEEE

```
ISEGEN()
01: last_best_C ⇐ C
02: loop (until exit condition)
03:     best_C ⇐ last_best_C
04:     while (there exists unmarked node in DFG)
05:         foreach (unmarked node n)
06:             Calculate Gain Function, $M_{toggle}(n, best\_C)$
07:         endfor
08:         best_node ⇐ Node with maximum Gain
09:         Toggle and Mark best_node
10:         Get impact of toggling best_node w.r.t. best_C
11:         if (toggling best_node satisfies constraints)
12:             Update best_C from toggling best_node
13:             Calculate $M(best\_C)$
14:         endif
15:     endwhile
16:     if ($M(best\_C) > M(last\_best\_C)$)
17:         last_best_C ⇐ best_C
18:         Unmark all nodes
19:     endif
20: endloop
21: C ⇐ last_best_C
```

**Figure 2.** *The ISEGEN Algorithm*

partition map to an ISE to be executed on an AFU while those belonging to the software partition individually execute on the processor core. Our approach considers the basic blocks in an application based on their speedup potential — a function of its execution frequency and estimated gain from mapping all its nodes to hardware — and performs up to $N_{ISE}$ successive bi-partitions into hardware and software within a basic block. After an ISE is found in a basic block, the speedup potential of the block is updated considering the remaining nodes.

We borrow the idea from *Kernighan-Lin* min-cut partitioning heuristic to steer **toggling** of nodes in the DFG between software (*S*) and hardware (*H*) based on a gain function that captures the designer's objective. The effectiveness of the K-L heuristic lies in its ability to overcome many local maxima without using unnecessary moves.

### 4.1 Modified Kernighan-Lin Algorithm

The ISEGEN algorithm that essentially performs a bi-partitioning of a DFG into *S* and *H* is depicted in Figure 2. This is an iterative improvement algorithm that starts with all nodes in software and tries to toggle each unmarked node, $n$, in the graph from *S* to *H* or *H* to *S* in every iteration. Within each iteration of ISEGEN (line 02 to line 20), $last\_best\_C$ retains the best cut found so far with the help of $best\_C$, that maintains the intermediate best cuts. Initially, the cut $C$ points to a configuration where all nodes belong to software and this configuration is passed down to $best\_C$. The decision to toggle $n$ with respect to $best\_C$ is based on a gain function, $M_{toggle}()$. The gain function is evaluated for each node (line 05) and the node with the best gain, $best\_node$ (obtained in line 08) is then toggled and marked (line 09). Note that the chosen cut at this point may be violating input/output constraints and convexity constraints. In other words, we allow a cut to be illegal giving it an opportunity to eventually grow into a valid cut.

If both convexity and I/O constraints are satisfied (line 11), $best\_C$ is updated through removal of $best\_node$ from the cut or its addition to the cut depending on whether $best\_node$ has toggled from *H* to *S* or *S* to *H* respectively. The speedup estimate $M()$ determines whether $best\_C$ should override $last\_best\_C$ (line 17). This process is carried on till no more unmarked nodes are left. In general, we found experimentally that 5 passes are enough for successive improvement of the solution. Therefore, the exit condition in the outermost loop is set to 5 times or lower when there is no improvement in the merit of the solution across successive iterations. The best cut ($last\_best\_C$) is stored back in $C$ that further acts as a starting point for the next bi-partitioning of the DFG.

### 4.2 Gain Function

The gain function $M_{toggle}$ for toggling a node $n$ with respect to a cut $C$ is a linear weighted sum of the following 5 components that act as control parameters for the algorithm:

- **Merit Function (Speedup Estimate)**: Let $C'$ be the new cut after addition or removal of the node $n$ from the cut $C$ as $n$ toggles from *S* to *H* or *H* to *S* respectively.

$$mrt = \begin{cases} M(C'), & \text{if } C' \text{ obeys convexity constraint,} \\ -\infty, & \text{if } C' \text{ violates convexity constraint.} \end{cases}$$

- **Input Output violation penalty**: A heavy penalty is applied with the help of a large factor if input-output port constraints are violated.

$$iop = \left( \left( I_{\text{ISE}}(C') - N_{\text{in}} \right) + \left( O_{\text{ISE}}(C') - N_{\text{out}} \right) \right),$$

- **Convexity Constraints**: Addition of a node to a cut is favored when its neighbors are already in the cut while a node already in the cut is not easily removed from the cut. Let $num\_nbrs\_in\_cut(n, C)$ be the number of neighbors of $n$ in $C$.

$$cnv = \begin{cases} +num\_nbrs\_in\_cut(n, C), & \text{if } n \text{ is in } S, \\ -num\_nbrs\_in\_cut(n, C), & \text{if } n \text{ is in } H. \end{cases}$$

- **Large Cut**: A cut is allowed to grow in regions where growth potential is higher. The external input and external output nodes act as barriers beyond which a cut cannot grow. Since we do not allow memory access from AFUs, memory operations are also barriers for cut growth. Let $d\_to\_bars\_up(n)$ be the minimum distance of $n$ from the barriers in the upward direction and let $d\_to\_bars\_down(n)$ be the minimum distance of $n$ from the barriers in the downward direction.

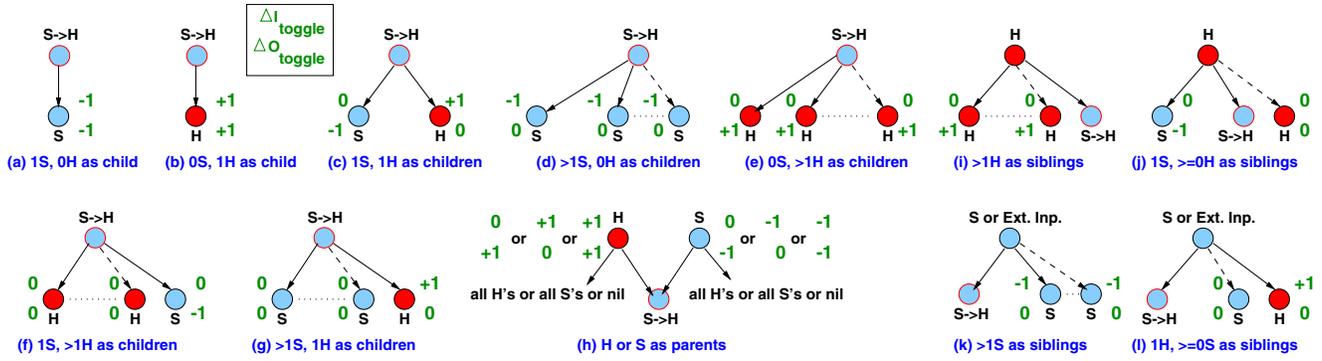

**Figure 3. Basic Rules to project the effect of toggling a node from S to H to its parents (h), children (a-g) and siblings (i-l). The ($\Delta I_{\text{toggle}}, \Delta O_{\text{toggle}}$) pair shown associated with each node that gets affected.**

$$cgp = \begin{cases} +|d\_to\_bars\_up(n) - d\_to\_bars\_down(n)|, \\ \quad\text{if } n \text{ is in } S, \\ -|d\_to\_bars\_up(n) - d\_to\_bars\_down(n)|, \\ \quad\text{if } n \text{ is in } H. \end{cases}$$

We employ a directional growth strategy where nodes closer to the barrier (that have higher potential for cut growth) are consistently favored for inclusion in hardware; this strategy implicitly favors reusability of the cut without losing the benefit of having large cut as a solution.

- **Independent Cuts**: It is quite possible that the best cut is actually a combination of 2 or 3 large connected subgraphs and not necessarily the largest connected subgraph. So, ISE exploration needs to expand not only in the vertical direction favoring large cuts but also in the horizontal direction. Let $CS(G)$ be the independently connected subgraphs in the DFG $G$ excluding the connected subgraph containing $n$.

$$idc = \begin{cases} +max_{cs \in CS(G)} CP\_lat(cs), \\ \quad\text{if } n \text{ is in } H, \\ 0, \text{if } n \text{ is in } S. \end{cases}$$

where $CP\_lat(cs)$ is the sum of the hardware latencies along the critical path of the independently connected subgraph, $cs$. Using this component, the nodes already in $H$ are allowed to move back into $S$ to favor the growth of other potentially large subgraphs.

We now express $M_{\text{toggle}}(n)$ with respect to the current cut $C$ as follows:

$$\alpha_1 \cdot mrt - \alpha_2 \cdot iop + \alpha_3 \cdot cnv + \alpha_4 \cdot cgp + \alpha_5 \cdot idc$$

The weights $\alpha_1$, $\alpha_2$, $\alpha_3$, $\alpha_4$ and $\alpha_5$ have been determined experimentally. We show in [9] that by maintaining appropriate data structures, the worst-case running time of ISEGEN can be restricted to $O(|V| \cdot |E|)$.

### 4.3 Impact of Toggling a Node

The runtime complexity of $M_{\text{toggle}}$ is significantly reduced by trading the majority of computations into appropriately evaluating the impact of toggling a node (line 10 in Figure 2). The number of inputs and the number of outputs of ISE at any stage of the partitioning process are given by $I_{\text{ISE}}$ and $O_{\text{ISE}}$ respectively. In order to quantify the impact of toggling a node, we introduce **addendums** $I_{\text{toggle}}$ and $O_{\text{toggle}}$ associated with every node. When a node is toggled, its addendums $I_{\text{toggle}}$ and $O_{\text{toggle}}$ are added to $I_{\text{ISE}}$ and $O_{\text{ISE}}$ respectively to get the new values of $I_{\text{ISE}}$ and $O_{\text{ISE}}$. Initially, all nodes are in $S$ and therefore $I_{\text{ISE}} = O_{\text{ISE}} = 0$ and $I_{\text{toggle}}$ and $O_{\text{toggle}}$ equal the number of inputs and number of outputs respectively of the corresponding node. It is easy to show that when a node is toggled (say, from $S$ to $H$), $I_{\text{toggle}}$ and $O_{\text{toggle}}$ of only its neighbors (parents, children and siblings) get affected. After toggling from $S$ to $H$, $I_{\text{toggle}}$ and $O_{\text{toggle}}$ of the node reverse in sign so that the changes to $I_{\text{ISE}}$ and $O_{\text{ISE}}$ will be undone if the same node toggles back to $S$. The impact of toggling a node (node 3) on itself and other nodes is illustrated in Figure 5.

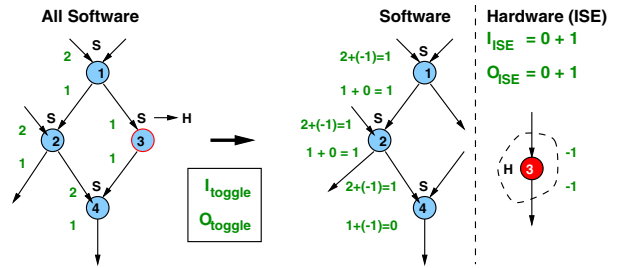

**Figure 5. Instance of an Instruction-level Hardware-Software Partitioning**

We developed a comprehensive set of rules to capture the effect of toggling a node that is pictorially presented in Figure 3 wherein a toggle of a node from $S$ to $H$ is shown as $S \rightarrow H$. The changes in $I_{\text{toggle}}$ and $O_{\text{toggle}}$ values are represented as $\Delta I_{\text{toggle}}$ and $\Delta O_{\text{toggle}}$ respectively such that the new values of $I_{\text{toggle}}$ and $O_{\text{toggle}}$ for the affected nodes are computed as $(I_{\text{toggle}} + \Delta I_{\text{toggle}})$ and $(O_{\text{toggle}} + \Delta O_{\text{toggle}})$ respectively. These rules can be empirically verified to work on any DFG (for example, Figure 5). An additional rule is that a toggle of a node from H to S negates the effect of its toggling from S, i.e., the rules in Figure 3 can be applied for toggling from H to S with the sign reversed for $\Delta I_{\text{toggle}}$ and $\Delta O_{\text{toggle}}$. The proofs of correctness for all the rules have been omitted for the sake of brevity and presented in [9]. The impact of toggling a node also involves maintenance of appropriate data structures for fast evaluation of M() and convexity violation [9].

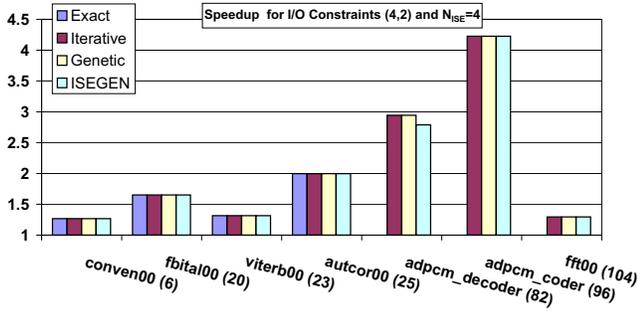 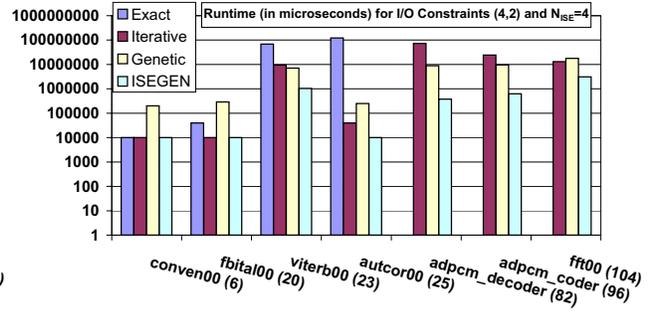

**Figure 4. Comparison of Speedup and Runtime with number of AFUs = 4 and I/O constraints: (4,2)**

## 5 Experimental Results

We define the merit function as: $\text{M}(C) = \lambda_{\text{sw}}(C) - \lambda_{\text{hw}}(C)$, where $\lambda_{\text{sw}}(C)$ is the software latency of $C$ estimated by summing the latencies of the nodes in $C$; $\lambda_{\text{hw}}(C)$ is the hardware latency of $C$ estimated from the critical path in $C$. The hardware latency for each instruction was obtained by synthesizing the constituent arithmetic and logic operators on a common $0.18\mu m$ CMOS technology and then normalized to the delay of a 32-bit *multiply-accumulate (MAC)*.

We integrated ISEGEN in the MachSUIF framework [12] and evaluated overall speedup for the entire application using all the generated cuts as follows:

$$\frac{\lambda_{\text{overall}}}{\lambda_{\text{overall}} - \sum_C N_C \cdot \text{M}(C)}$$

The variable, $\lambda_{\text{overall}}$ encapsulates the overall execution latency of the application i.e., when the application entirely runs on software, and $N_C$ is the execution frequency of $C$. Note that, in this work, we do not consider memory operations for inclusion into a cut.

To evaluate the efficacy of our ISEGEN approach, we chose benchmarks from diverse application domains in EEMBC (*autcor00*, *viterbi00*, *conven00*, *fft00* and *fbital00*) and MediaBench (*adpcm coder* and *adpcm decoder*) suites. In addition, we chose a cryptographic application viz. $AES$. Our baseline architecture is a simple RISC machine and we allow up to 4 AFUs (or ISEs) to be added. Keeping the I/O constraints fixed at $(4, 2)$, we study the overall speedup of applications obtained over execution on the core processor and the time taken to generate ISEs (or runtime) on Sun Ultra-5. We compare the quality of our results with the best known algorithms for ISE generation. The optimal algorithms for ISE generation [3] come in two flavors: Exact multiple-cut identification (or **Exact** in short) and Iterative exact single-cut identification (or **Iterative**), both of which employ exhaustive search with pruning. For applications having large basic blocks, we chose a genetic formulation [4] for comparing our results.

We associate with each benchmark the maximum number of nodes in its critical basic block (shown in parentheses) and arrange them in increasing order. It is evident from the first plot of Figure 4 that ISEGEN matches the solution quality of Exact, Iterative and Genetic algorithms. Note that because of effective pruning, Exact is able to handle up to 25 nodes and Iterative is able to handle up to 104 nodes in the selected benchmarks. As shown in the second plot of Figure 4, ISEGEN runs up to $29 \times$ faster than the genetic approach with the generated ISEs having quality comparable with the optimal solution in terms of overall speedup. We observed that some of the ISEs identified by the optimal algorithms are independent subgraphs and therefore an ISE identification algorithm should not be restricted to identify only connected subgraphs.

$AES$ is a cryptographic benchmark with a large DFG; its critical basic block contains 696 nodes with a symmetric structure. Since the optimal algorithms (Exact and Iterative) could not run on such a large application, we chose the genetic solution (that also matches the optimal solution in smaller benchmarks) for comparing our results. Because of its non-exponential complexity, ISEGEN easily handles large DFGs. We deliberately chose $AES$ to demonstrate the efficacy of our ISEGEN approach in matching expert design quality. We increased the maximum number of AFUs from 1 to 4, and studied the speedup over execution on the processor core as shown in Figure 6.

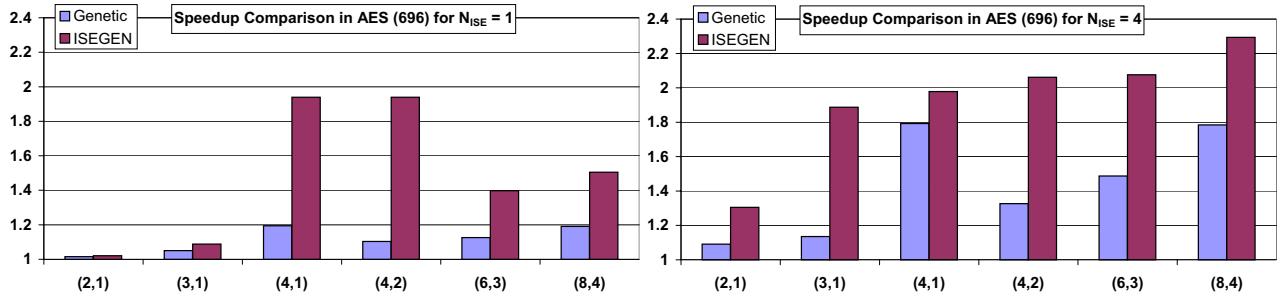

**Figure 6. Comparison of Speedup on AES with varying number of AFUs**

On average, ISEGEN obtains 35% more speedup than the genetic solution by effectively exploiting the regularity in the data flow graph of AES. Figure 7 shows how the structure yielded multiple instances of the same cut thereby exposing the regularity in the application. Since AES has a large number of nodes, it is intuitive to expect an increase in speedup by increasing the allowed number of AFUs and I/O constraints. However, it is interesting to note that contrarily to our expectation, for a smaller number of allowed AFUs (= 1), the speedup could not scale with relaxing I/O constraints (as shown in the first plot of Figure 6). The reason is clear from the plot of Figure 7. It shows that there are 12 instances of the first cut for the I/O constraint of (4, 1) (or (4, 2)), while there are only 4 instances for the I/O constraint of (6, 3). As is evident from the first plot of Figure 6, the 12 instances generated for (4, 1) cover the DFG better than the 4 instances generated for (6, 3). However, with increase in the allowed number of AFUs, the speedup begins to scale with increasing I/O constraints (as shown in the last plot of Figure 6). Therefore, our ISEGEN not only generates ISEs resulting in high speedup but also exploits their reusability by producing all the instances in the DFG (as shown in Figure 7). Thus, the solutions generated by ISEGEN are indeed close to those generated by an expert designer.

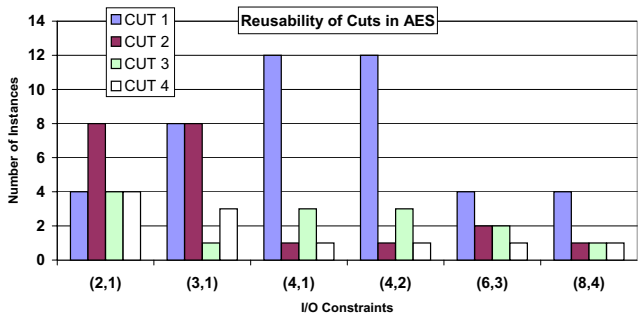

**Figure 7. Study of Reusability of ISEs on AES with varying number of AFUs**

## 6 Conclusions

The hardware-software partitioning problem when applied at the instruction-level granularity constitutes the problem of ISE generation. The contributions presented in this paper are as follows. First, we clearly identified the properties of ISEs that are of interest to an expert designer. Second, we adapted a well-known Kernighan-Lin heuristic to perform ISE generation with a low computational complexity. Finally, we show that our ISEGEN approach produces high-quality ISEs — close to those sought after by an expert designer. Furthermore, ISEGEN runs up to $29\times$ faster than a previous genetic approach and generates solutions comparable with the optimal ISE generation approaches. Our future work will focus on the deployment of ISEs in a real system and evaluating the impact of ISEs on code size and energy reduction.

## References

[1] F. Vahid and T. D. Le. Extending the Kernighan/Lin Heuristic for Hardware and Software Functional Partitioning. In *Kluwer Journal on Design Automation of Embedded Systems*, 1997.

[2] C. M. Fiduccia and R. M. Mattheyses. A Linear-time Heuristic for Improving Network Partitions. In *Proc. of DAC*, 1982.

[3] K. Atasu, L. Pozzi and P. Ienne. Automatic Application-Specific Instruction-Set Extensions under Microarchitectural Constraints. In *Proc. of DAC*, 2003.

[4] P. Biswas, V. Choudhary, K. Atasu, L. Pozzi, P. Ienne and N. Dutt. Introduction of Local Memory Elements in Instruction Set Extensions. In *Proc. of DAC*, 2004.

[5] N. Clark, H. Zhong and S. Mahlke. Processor Acceleration through Automated Instruction Set Customization. In *Proc. of MICRO*, 2003.

[6] P. Yu and T. Mitra. Scalable Custom Instructions Identification for Instruction-Set Extensible Processors. In *Proc. of CASES*, 2004.

[7] F. Sun, S. Ravi, A. Raghunathan and N. K. Jha. Synthesis of Custom Processors based on Extensible Platforms. In *Proc. of ICCAD*, 2002.

[8] M. Arnold and H. Corporaal. Designing Domain-specific Processors. In *Proc. of CODES*, 2001.

[9] P. Biswas, S. Banerjee, N. Dutt, L. Pozzi and P. Ienne. ISEGEN: Adapting Kernighan-Lin Min-Cut Heuristic for Generation of Instruction Set Extensions. CECS, UC Irvine, Technical Report CECS-TR-04-21.

[10] C. Alippi, W. Fornaciari, L. Pozzi and M. Sami. A DAG based Design Approach for Reconfigurable VLIW Processors. In *Proc. of DATE*, 1999.

[11] R. Razdan and M. D. Smith. A High-performance Microarchitecture with Hardware-programmable Functional Units. In *Proc. of MICRO*, 1994.

[12] Machine SUIF. http://www.eecs.harvard.edu/hube/software/software.html.